\input harvmac
\sequentialequations
%\draftmode
\overfullrule=0pt
\def\bigZ{Z\!\!\!Z}
\def\twelfth{{\textstyle{1\over12}}}
\def\utilde{\tilde u}
\def\Pinfty{{\cal P}_{\sst\infty}}
\def\Atot{{\A_{\rm tot}}}
\def\tot{{\rm tot}}
\def\uA{\,\lower 1.2ex\hbox{$\sim$}\mkern-13.5mu A}
\def\bigL{{\bf L}}

\def\zero{{\scriptscriptstyle(0)}}
\font\authorfont=cmcsc10 \ifx\answ\bigans\else scaled\magstep1\fi
%\divide\baselineskip by 7
%\multiply\baselineskip by 6
\divide\baselineskip by 10
\multiply\baselineskip by 9
%\advance\topskip by -1in\relax
{\divide\baselineskip by 4
\multiply\baselineskip by 4
\def\prenomat{\matrix{\hbox{hep-th/9612231}&\cr \qquad
  \hbox{NO-EL/96}  
&\cr}}
\Title{$\prenomat$}{\vbox{\centerline{On Mass-Deformed $N=4$
Supersymmetric Yang-Mills Theory
}}}
\centerline{\authorfont Nicholas Dorey}
\bigskip
\centerline{\sl Physics Department, University of Wales Swansea}
\centerline{\sl Swansea SA2$\,$8PP UK $\quad$ \tt n.dorey@swansea.ac.uk}
\bigskip
\centerline{\authorfont Valentin V. Khoze}
\bigskip
\centerline{\sl Department of Physics, Centre for Particle Theory, 
University of Durham}
\centerline{\sl Durham DH1$\,$3LE UK $\quad$ \tt valya.khoze@durham.ac.uk}
\bigskip
\centerline{and}
\bigskip
\centerline{\authorfont Michael P. Mattis}
\bigskip
\centerline{\sl Theoretical Division T-8, Los Alamos National Laboratory}
\centerline{\sl Los Alamos, NM 87545 USA$\quad$ \tt mattis@pion.lanl.gov}
\vskip .3in
\def\hf{{\textstyle{1\over2}}}
\def\quarter{{\textstyle{1\over4}}}
\noindent
We construct the $n$-instanton action for the above model with gauge group
$SU(2)$, as a function of
the collective coordinates of the general self-dual configurations of
Atiyah, Drinfeld, Hitchin and Manin (ADHM). We calculate the quantum
modulus $u = \langle\Tr\uA^2\rangle$ at the 1-instanton level, and find
a discrepancy with Seiberg and Witten's proposed exact solution. As in
 related models ($N=2,\ N_F=3 \hbox{ or } 4$), 
this discrepancy may be resolved by 
modifying their proposed
 relation between $\tilde u$ (the parameter in the elliptic curve) and
$u$. 
%\looseness=-1
%\relax
\vskip .1in
\Date{\bf December 1996 } %replace this line
%by \draft  for preliminary versions or specify \draftmode at some point
\vfil\break
}

\lref\Seiberg{ N. Seiberg, Phys. Lett. B206 (1988) 75. }
\lref\ADHM{ M.  Atiyah, V.  Drinfeld, N.  Hitchin and
Yu.~Manin, Phys. Lett. A65 (1978) 185. }
\lref\Ferrari{F.Ferrari, private communication, and in preparation.}
\lref\ItoSas{K. Ito and N. Sasakura, \it One-instanton calculations
in $N=2$ $SU(N_c)$ supersymmetric QCD\rm, hep-th/9609104.}
\lref\CGTone{ E. Corrigan, P. Goddard and S. Templeton,
Nucl. Phys. B151 (1979) 93; \hfil\break
   E. Corrigan, D. Fairlie, P. Goddard and S. Templeton,
    Nucl. Phys. B140 (1978) 31.}
\lref\Fucito{F. Fucito and G. Travaglini, 
{\it Instanton calculus and nonperturbative relations
 in $N=2$ supersymmetric gauge theories}, 
 ROM2F-96-32, hep-th/9605215.}
\lref\dkmone{N. Dorey, V.V. Khoze and M.P. Mattis, \it Multi-instanton
calculus in $N=2$ supersymmetric gauge theory\rm, hep-th/9603136,
Phys.~Rev.~D54 (1996) 2921.}
\lref\dkmseven{N. Dorey, V.V. Khoze and M.P. Mattis, in preparation.}
\lref\dkmfive{N. Dorey, V.V. Khoze and M.P. Mattis, \it On $N=2$
supersymmetric QCD with $4$ flavors\rm,  hep-th/9611016.}
\lref\dkmtwo{N. Dorey, V.V. Khoze and M.P. Mattis, 
\it Multi-instanton check of the relation between the prepotential  
${\cal F}$ 
and the modulus $u$ in $N=2$ SUSY Yang-Mills theory\rm,
hep-th/9606199, Phys.~Lett.~B389 (in press, 26 December 1996).}
\lref\dkmthree{N. Dorey, V.V. Khoze and M.P. Mattis, 
\it A two-instanton test of the exact solution of $N=2$ supersymmetric  
QCD\rm,
hep-th/9607066, Phys.~Lett.~B388 (1996) 324.}
\lref\dkmfour{N. Dorey, V.V. Khoze and M.P. Mattis, \it Multi-instanton
calculus in $N=2$ supersymmetric gauge theory.
II. Coupling to matter\rm, hep-th/9607202, Phys.~Rev.~D54 (1996) 7832.}
\lref\HS{T. Harano and M. Sato, \it Multi-instanton
calculus versus exact results in $N=2$ supersymmetric QCD\rm,
hep-th/9608060}
\lref\Aoyama{H. Aoyama, T. Harano, M. Sato and S.Wada,
\it   Multi-instanton calculus in $N=2$ supersymmetric QCD\rm,
hep-th/9607076.}   
\lref\SWone{N. Seiberg and E. Witten, 
{\it Electric-magnetic duality, monopole
condensation, and confinement in $N=2$ supersymmetric Yang-Mills  
theory}, 
Nucl. Phys. B426 (1994) 19, (E) B430 (1994) 485  hep-th/9407087}
\lref\SWtwo{
N. Seiberg and E. Witten, 
{\it Monopoles, duality and chiral symmetry breaking
in $N=2$ supersymmetric QCD}, 
Nucl. Phys B431 (1994) 484 ,  hep-th/9408099}
\lref\BMSTY{T. Eguchi and S.K. Yang, 
{\it Prepotentials of $N=2$ supersymmetric gauge theories 
and soliton equations}, hep-th/9510183;
\hfil\break
G. Bonelli and M. Matone, 
    Phys. Rev. Lett. 76 (1996) 4107, hep-th/9602174; \hfil\break
 J. Sonnenschein, S. Theisen and S. Yankielowicz,
 Phys. Lett. 367B (1996) 145, hep-th/9510129.}
\lref\MAtone{
M. Matone, {\it Instantons and recursion relations in $N=2$ SUSY gauge  
theory},
 Phys. Lett. B357 (1995) 342,    hep-th/9506102.  }
\lref\FPone{ D. Finnell and P. Pouliot,
{\it Instanton calculations versus exact results in $4$ dimensional 
SUSY gauge theories},
Nucl. Phys. B453 (95) 225, hep-th/9503115. }
\lref\tHooft{G. 't Hooft, Phys. Rev. D14 (1976) 3432; ibid.
D18 (1978) 2199.}
\def\frac#1#2{{ {#1}\over{#2}}}

\def\inst{{\rm inst}}

\def\trtwo{\tr^{}_2\,}

\def\wbar{\bar w}

\def\abar{\bar a}

\def\dalpha{{\dot\alpha}}

\def\sst{\scriptscriptstyle}

\def\PV{{\sst\rm PV}}

\def\F{{\cal F}}

\def\A{{\cal A}}
\def\susy{supersymmetry}

\def\cl{{\,\rm cl}}
\def\lambdabar{\bar\lambda}
\def\R{{R}}
\def\psibar{\bar\psi}
\def\sqrtwo{\sqrt{2}\,}
\def\etabar{\bar\eta}

\def\Qbar{\bar Q}
\def\susic{supersymmetric}

\def\vhiggs{{\rm v}}

\def\vhiggsa{{\cal A}_{\sst00}}
\def\vbarhiggs{\bar{\rm v}}

\def\C{{\cal C}}

\def\new{{\scriptscriptstyle\rm new}}

\def\uX{\,\lower 1.2ex\hbox{$\sim$}\mkern-13.5mu X}
\def\uQ{\,\lower 1.2ex\hbox{$\sim$}\mkern-13.5mu Q}
\def\uQtilde{\,\lower 1.2ex\hbox{$\sim$}\mkern-13.5mu \tilde Q}
\def\uD{\,\lower 1.2ex\hbox{$\sim$}\mkern-13.5mu {\rm D}}

\def\uF{\,\lower 1.2ex\hbox{$\sim$}\mkern-13.5mu F}
\def\uW{\,\lower 1.2ex\hbox{$\sim$}\mkern-13.5mu W}
\def\uWbar{\,\lower 1.2ex\hbox{$\sim$}\mkern-13.5mu {\overline W}}
\def\uPhibar{\,\lower 1.2ex\hbox{$\sim$}\mkern-13.5mu {\overline \Phi}}

\def\uAbar{{\uA}^\dagger}

\def\uV{\,\lower 1.2ex\hbox{$\sim$}\mkern-13.5mu V}
\def\uv{\,\lower 1.0ex\hbox{$\scriptstyle\sim$}\mkern-11.0mu v}
\def\uPsi{\,\lower 1.2ex\hbox{$\sim$}\mkern-13.5mu \Psi}
\def\uPhi{\,\lower 1.2ex\hbox{$\sim$}\mkern-13.5mu \Phi}
\def\uchi{\,\lower 1.5ex\hbox{$\sim$}\mkern-13.5mu \chi}
\def\uchitilde{\,\lower 1.5ex\hbox{$\sim$}\mkern-13.5mu \tilde\chi}
\def\Psibar{\bar\Psi}
\def\uPsibar{\,\lower 1.2ex\hbox{$\sim$}\mkern-13.5mu \Psibar}
\def\upsi{\,\lower 1.5ex\hbox{$\sim$}\mkern-13.5mu \psi}
\def\uq{\,\lower 1.5ex\hbox{$\sim$}\mkern-13.5mu q}
\def\uqtilde{\,\lower 1.5ex\hbox{$\sim$}\mkern-13.5mu \tilde q}
\def\psibar{\bar\psi}
\def\upsibar{\,\lower 1.5ex\hbox{$\sim$}\mkern-13.5mu \psibar}
\def\upsibarzero{\,\lower 1.5ex\hbox{$\sim$}\mkern-13.5mu \psibar^\zero}
\def\ulambda{\,\lower 1.2ex\hbox{$\sim$}\mkern-13.5mu \lambda}
\def\ulambdabar{\,\lower 1.2ex\hbox{$\sim$}\mkern-13.5mu \lambdabar}
\def\ulambdabarzero{\,\lower 1.2ex\hbox{$\sim$}\mkern-13.5mu \lambdabar^\zero}
\def\ulambdabarnew{\,\lower 1.2ex\hbox{$\sim$}\mkern-13.5mu \lambdabar^\new}
\def\D{{\cal D}}
\def\M{{\cal M}}
\def\N{{\cal N}}
\def\Dslash{\,\,{\raise.15ex\hbox{/}\mkern-12mu \D}}
\def\Dbarslash{\,\,{\raise.15ex\hbox{/}\mkern-12mu {\bar\D}}}
\def\delslash{\,\,{\raise.15ex\hbox{/}\mkern-9mu \partial}}
\def\delbarslash{\,\,{\raise.15ex\hbox{/}\mkern-9mu {\bar\partial}}}
\def\L{{\cal L}}
\def\hf{{\textstyle{1\over2}}}
\def\quarter{{\textstyle{1\over4}}}
\def\eighth{{\textstyle{1\over8}}}

\def\xibar{\bar\xi}

\def\uAcl{\,\lower 1.2ex\hbox{$\sim$}\mkern-13.5mu A^{}_{\cl}}
\def\uAbarcl{\,\lower 1.2ex\hbox{$\sim$}\mkern-13.5mu A_{\cl}^\dagger}

%\newsec{Introduction}
\bf 1. Introduction. \rm
This Letter continues the program developed in 
Refs.~\refs{\FPone\dkmone\dkmtwo\dkmthree\dkmfour\dkmfive\Aoyama-\HS},
in which Seiberg and Witten's proposed exact solutions to a variety of
$N=2$ \susic\ $SU(2)$ gauge theories \refs{\SWone,\SWtwo} are tested
against first-principles (multi-)instanton calculations in the semiclassical
regime. The case of pure $N=2$ supersymmetric Yang-Mills (SYM) theory is
treated in Refs.~\refs{\SWone,\FPone-\dkmtwo}, whereas in 
Refs.~\refs{\SWtwo,\dkmthree-\HS} the model is augmented by $N_F$ flavors
of ``quark hypermultiplets'' that transform in the fundamental representation
of the gauge group. Also treated in Ref.~\SWtwo, and the topic of the
present Letter, is the case of a single massive
hypermultiplet in the \it adjoint
\rm representation of $SU(2)$. (For more than one such hypermultiplet,
just as for $N_F>4$ fundamental hypermultiplets, the $\beta$-function
is positive, and the microscopic theory no longer makes sense.) 
The term ``mass-deformed $N=4$ SYM theory'' is used to describe this
model, since in the limit that the hypermultiplet is massless, the 
$N=2$ algebra is enlarged to $N=4,$ and the $SU(2)_R$ symmetry that
acts on the supercharges is enlarged to $SU(4)_R.$  In this limit the
number of unbroken fermion zero modes of the (multi-)instanton doubles
from four to eight, guaranteeing that the low-energy dynamics receives
neither instanton nor perturbative corrections \Seiberg. But for nonzero
mass these extra zero modes are lifted and such corrections do occur,
as discussed in detail below.

The mass-deformed $N=4$
model  shares a number of interesting features with another model
discussed in \refs{\SWtwo,\dkmfive},
 namely the $N_F=4$ model. First, both are finite theories.
Second, when the bare hypermultiplet
 masses vanish, both models are conformally invariant.
The low-energy $U(1)$ dynamics along the Coulomb branches
 are then simply described by a  free field theory, that
of a massless uncharged $N=4$ and $N=2$ gauge superfield, 
respectively. Third, the BPS dyon spectrum in the massless models is generally
thought to be $SL(2,\bigZ)$ invariant.\foot{This is apparently not the
case when masses are turned on \Ferrari.} 
Fourth, even in the massive case, the elliptic curves 
governing the proposed exact solutions are built from modular forms in
the complexified coupling $\tau$;
%As argued in Sec.~16 of \SWtwo,
%this implies that there
%are an infinite number of distinct  values of $\tau$ about which the
%theory admits a weakly-coupled description in the appropriate dual variables.
%As for the BPS dyon spectrum in these finite theories, it now appears that
%it is \it not \rm $SL(2,\bigZ)$ invariant at a given point on the
%moduli space so long as masses
%are turned on \Ferrari\ or for gauge groups higher than $SU(2)$ 
%\Cederwall.
in fact  the  curve for the former model
may be obtained from the latter model, by restricting to the 
special 
case that the four ``quark'' masses are $\{m/2\,,\,m/2\,,\,0,0\}$ \SWtwo. 
There is however a key difference:  in this mapping the
$n$-instanton sector of the former corresponds to the $2n$-instanton
sector of the latter. This is related to the fact that, 
unlike the mass-deformed $N=4$ model, the
$N_F=1,2,3,4$ models have a $\bigZ_2$ symmetry forbidding
odd-instanton contributions when at least one ``quark'' is massless
\SWtwo.  It is therefore not surprising that, below,
we will uncover a discrepancy at the 1-instanton level
with Seiberg and Witten's prediction for the quantum modulus
$u = \langle\Tr\uA^2\rangle$, analogous to a 2-instanton discrepancy
in the $N_F=3$ \refs{\Aoyama,\HS} and $N_F=4$ models \refs{\dkmfour,\dkmfive}.
(Similar \hbox{1-instanton} discrepancies have been claimed 
for $SU(3)$ gauge theory for $N_F=4,6$  \ItoSas.) Another key
difference between the two 
models is that, in the massless limit, the low-energy
dynamics of the 
$N_F=4$ theory \it does \rm receive interesting finite numerical 
perturbative and instanton renormalizations \refs{\dkmfour,\dkmfive}.

The particle content of the  mass-deformed $N=4$ model is most easily 
understood in the familiar language of $N=1$ superfields. In these terms
the $N=2$ SYM Lagrangian is built from a gauge superfield\foot{We 
use undertwiddling as a shorthand for adjoint $SU(2)$ fields; thus
${\uX}=\sum_{a=1,2,3}\,X^a\tau^a/2$, where $\tau^a$ are Pauli matrices.}
$\uW^\alpha=(\uv_m,\ulambda\,)$ coupled
to a  chiral superfield $\uPhi=(\uA,\upsi\,)$.
The adjoint $N=2$ hypermultiplet is described by a pair of $N=1$ 
chiral superfields $\uQ=(\uq,\uchi\,)$ and $\uQtilde=(\uqtilde,\uchitilde\,)$.
These couple to $\uW^\alpha$ in the standard way, and to $\uPhi$
via the superpotential
\eqn\superptl{{\cal L}_{\rm int}\ =\ \trtwo\big(
4\sqrtwo i\uQtilde\uPhi\uQ+2im\uQtilde\uQ\big)\Big|_{\theta^2}\ +\ {\rm H.c.}}
The coefficient of $\uQtilde\uPhi\uQ$ is  equated to that of the other
Yukawa terms in the component Lagrangian, as they transform
 among themselves under
$SU(4)_R$ symmetry. In turn, the coefficient of the mass term in \superptl\
is fixed by Seiberg and Witten's convention \SWtwo\ that the semiclassical
singularity lies at $u\simeq m^2/4$ where $u=\hf\vhiggs^2+$ (instanton
corrections). Indeed, with Higgs VEVs $\vev{\uA\,}=\vhiggs\tau^3/2$ and
$\vev{\uq\,}=\vev{\uqtilde\,}=0$, a tree-level diagonalization of
\superptl\ gives mass eigenvalues $\{m\,,\,m+\sqrtwo \vhiggs\,,\,
m-\sqrtwo \vhiggs
\}$ with zeroes at $\sqrtwo \vhiggs=\pm m$ in accord with this convention.

This Letter is organized as follows. In Sec.~2 we construct the
$n$-instanton action for the mass-deformed $N=4$ model, starting
with the known results for $N=2$ SYM theory \dkmone. In Sec.~3 we focus
on the 1-instanton sector, and calculate from first principles
the prepotential $\F$, and the
Higgs condensate $u$. In Sec.~4 we extract $u$ instead from
the predictions of Seiberg and Witten \SWtwo, and uncover a discrepancy
with the 1-instanton result. This discrepancy is resolved,
and the 2-instanton sector \dkmseven\ is qualitatively discussed, in Sec.~5.

%\newsec{Construction of the $n$-instanton action}
\bf 2. Construction of the $n$-instanton action. \rm
One of the chief results of \dkmone\ is the construction of the $n$-instanton
action for the case of pure $N=2$ SYM theory.\foot{This section relies heavily
on the formalism developed in Secs.~6-7 of \dkmone\ and
Secs.~2,3,5 of \dkmfour. In particular, see Sec.~6 of \dkmone\ for
a pedagogical introduction to Atiyah-Drinfeld-Hitchin-Manin (ADHM) 
theory \ADHM\ including our notational
conventions and references to the early literature. We set $g=1$ throughout
except, for clarity, in the instanton action $8\pi^2/g^2.$} 
The result is expressed in
terms of the $(n+1)\times n$ ADHM collective coordinate matrices $a,$
$\M$ \hbox{and $\N\,$:}
\eqn\bcanonical{
a_{\alpha\dalpha} =
\pmatrix{w_{1\alpha\dalpha}&\cdots&w_{n\alpha\dalpha}
\cr{}&{}&{}\cr
{}&a'_{\alpha\dalpha}&{}\cr{}&{}&{}}\ ,\ \
\M^\gamma = \pmatrix{\mu_1^\gamma&\cdots&\mu_n^\gamma
\cr{}&{}&{}\cr
{}&\M^{\prime\gamma}&{}\cr{}&{}&{}}
\ ,\ \
\N^\gamma = \pmatrix{\nu_1^\gamma&\cdots&\nu_n^\gamma
\cr{}&{}&{}\cr
{}&\N^{\prime\gamma}&{}\cr{}&{}&{}} }
In our conventions
the $n\times n$ submatrices $a',$ $\M'$ and $\N'$ are symmetric:
$a'=a^{\prime T},$
$\M'=\M^{\prime T},$ and
$\N'=\N^{\prime T}.$
The entries of $a$ are quaternion-valued, e.g., $w_{1\alpha\dalpha}=
w_{1m}\sigma^m_{\alpha\dalpha}$ where the $\sigma^m$ are the four spin
matrices $(1,i\vec\tau\,)$. 
In contrast, the entries of $\M$ and $\N$ are \hbox{2-component}
Weyl spinors. $\M$ and $\N$ parametrize the adjoint fermion zero modes
of the gaugino $\lambda$ and Higgsino
$\psi,$ respectively; together they form
an $SU(2)_R$ doublet. A drawback of the ADHM parametrization \bcanonical\
is that, for general $n>1,$ it is a highly overcomplete 
description of a supersymmetrized self-dual configuration
of topological number $n$. In order to reduce to $16n$ physical
degrees  of freedom (asymptotically,
$4n$ positions, $3n$ iso-orientations, and $n$
instanton scale sizes, plus superpartners),
one needs to impose not only the well-known constraints 
\refs{\ADHM,\CGTone}
\eqn\constraints{\abar a\ =\ (\abar a)^T\ ,\quad
\abar^{\dalpha\gamma}\M_\gamma\ =\ -\M^{\gamma T}a_\gamma
{}^\dalpha\ ,\quad
\abar^{\dalpha\gamma}\N_\gamma\ =\ -\N^{\gamma T}a_\gamma
{}^\dalpha\ ,}
but also further  ``gauge fixing conditions'' for the remaining
$O(n)$ redundancies as reviewed in detail in \dkmone.

Another 
key ingredient needed for the action is the self-adjoint linear operator
$\bigL,$ which is a bijection on the space of $n\times n$ antisymmetric
scalar-valued matrices. Explicitly,
if $\Omega$ is such a matrix, then $\bigL$ is defined as \dkmone
\eqn\bigLreally{\bigL\cdot\Omega\ =\ 
\hf\{\,\Omega\,,\,W\,\}\,-\,\hf\trtwo\big(
[\,\abar'\,,\,\Omega\,]a'-\abar'[\,a'\,,\,\Omega\,]\big)}
where $W$ is the symmetric  $n\times n$ matrix
$W_{kl}\,=\,\trtwo\wbar_kw_l\,.$
Important examples of such antisymmetric
matrices are $\Lambda,\Lambda_f,\A'$ and $\A'_f.$ 
The first two are defined by
\eqn\Lamsdef{\Lambda_{lk}\ =\ \wbar_l\vhiggsa w_k
-\wbar_k\vhiggsa w_l\ ,\quad
\Lambda_f\ =\ {1\over2\sqrtwo}\,
\big(\,\M^{\beta T}\N_\beta-\N^{\beta T}\M_\beta\,\big)\ ,}
where $\vhiggsa$ is the $SU(2)$-valued VEV,
\def\barvhiggsa{\bar{\cal A}_{\sst00}}
${\vhiggsa} = \textstyle{i\over2}\,\vhiggs\,\tau^3\,.$
In turn, $\A'$ and $\A'_f$ are defined implicitly as the solutions to
\eqn\Asdef{\bigL\cdot\A'\ =\ \Lambda\ ,\quad
\bigL\cdot\A_f'\ =\ \Lambda_f\ .}
%$\bigL\cdot\A' = \Lambda$  and $\bigL\cdot\A_f' = \Lambda_f$.
Whereas $\Lambda$ and $\A'$ are purely bosonic, $\Lambda_f$
and $\A'_f$ are fermion bilinears. Henceforth we will exhibit
 their two Grassmann arguments: $\Lambda_f=\Lambda_f(\M,\N)$
and $\A'_f=\A'_f(\M,\N).$ This will allow us to incorporate the
adjoint hypermultiplet in a parallel way.

In terms of these quantities, 
the $n$-instanton action for $N=2$ SYM theory reads  \dkmone:
\def\Lambdabar{\bar\Lambda}

\eqn\sinstfinal{\eqalign{S^0_{\rm inst}\  &=\
{8n\pi^2\over g^2}\ +\ 16\pi^2|\vhiggsa|^2\sum_{k=1}^n|w_k|^2
\ -\ 8\pi^2\,\Tr_n\big(\big[\A'+\A'_{f}(\M,\N)\big]\Lambdabar\big)
\cr&\ \ +\   
4\sqrtwo\pi^2\,\mu_k^\alpha\,\barvhiggsa{}_\alpha{}^\beta\,\nu_{k\beta}\ 
 .}}
A useful check of this expression is to show
that it is a supersymmetric invariant.
In fact, in Sec.~2 of \dkmfour, we demonstrated that the $N=2$ algebra can be
realized directly as an action on the overcomplete set of ADHM parameters
\bcanonical. Under an infinitesimal \susy\ transformation 
$\sum_{i=1,2}\,\xi_iQ_i+\xibar_i\Qbar_i\,$, they transform as:
\eqn\susytrans{\eqalign{\delta a_{\alpha\dalpha} \ &=\
 \xibar_{1\dalpha}\M_\alpha
+\xibar_{2\dalpha}\N_\alpha  \ ,\quad 
\delta\M_\gamma\ 
=\ -4ib\xi_{1\gamma}-2\sqrtwo\C_{\gamma\dalpha}(\M,\N;\vhiggsa)
\xibar_2^\dalpha\ ,\cr
\delta\N_\gamma\ &
=\ -4ib\xi_{2\gamma}+2\sqrtwo \C_{\gamma\dalpha}(\M,\N;\vhiggsa)
\xibar_1^\dalpha\ ,\quad
\delta\Atot(\M,\N;\vhiggsa)\ =\  0\  .}}
Here $\Atot(\M,\N;\vhiggsa)=\A'(\vhiggsa)+\A'_f(\M,\N)$, and 
\eqn\Cdef{\C(\M,\N;\vhiggsa)\ =\ \Bigg(
{\vhiggsa w_1\hbox{--}w_k\Atot{}_{k1}(\M,\N;\vhiggsa)\ \cdots\ \vhiggsa  
w_n\hbox{--}w_k\Atot{}_{kn}(\M,\N;\vhiggsa)
\atop
^{\phantom X}\big[\,\Atot(\M,\N;\vhiggsa)\,,\,a'\,\big]^{\phantom X}}\Bigg)\ .}
It is then easily verified that $\delta S^0_{\rm inst} =0$ as 
required \dkmfour.

We now describe how to augment the $N=2$ SYM expression \sinstfinal\ to
incorporate an adjoint hypermultiplet. We first discuss the mass
contribution $\L_{\rm mass}=2im\,\trtwo\uQtilde\uQ\big|_{\theta^2=0}^{}$ to the
superpotential \superptl. We associate to the hypermultiplet Higgsinos
$\uchi$ and $\uchitilde$, respectively, the Weyl-valued $(n+1)\times n$ 
collective coordinate matrices
\def\R{{\cal R}}
\def\Rtilde{\tilde{\cal R}}
\def\rhotilde{\tilde\rho}
\eqn\newcanon{\R^\gamma = \pmatrix{\rho_1^\gamma&\cdots&\rho_n^\gamma
\cr{}&{}&{}\cr
{}&\R^{\prime\gamma}&{}\cr{}&{}&{}}
\ ,\ \
\Rtilde^\gamma = \pmatrix{\rhotilde_1^\gamma&\cdots&\rhotilde_n^\gamma
\cr{}&{}&{}\cr
{}&\Rtilde^{\prime\gamma}&{}\cr{}&{}&{}} \ .}
Just like $\M$ and $\N,$ $\R$ and $\Rtilde$ satisfy $\R^{\prime T}=\R'$
and $\Rtilde^{\prime T}=\Rtilde'$ as well as 
constraints analogous to \constraints. Using
Corrigan's formula for the overlap of two adjoint zero modes (see Apps.~B-C of
\dkmone),  one finds 
\def\mass{{\rm mass}}
\eqn\Smassdef{S_\mass\ =\ \int d^4x\,\L_\mass\ =\ 2im\pi^2\,\Tr_n\,\Rtilde^T
(\Pinfty+1)\R}
where, 
in our conventions, $\Pinfty$ is the $(n+1)\times(n+1)$ matrix with a `1'
in 
the upper left-hand corner and zeroes elsewhere. (There are bosonic
mass terms too
but their effect on the semiclassical physics is
 down by one factor of the coupling as they require the
elimination of an auxiliary $F$ field.) In order to check that $S_\mass$ is 
invariant 
under $N=2$ \susy, one needs the transformation properties of $\R$ and
$\Rtilde.$ By manipulations identical to those used in Sec.~2 of \dkmfour\ to
derive the transformation laws \susytrans, one finds
\eqn\Rtrans{\delta\R_\gamma\ =\ 
-2\sqrtwo\C_{\gamma\dalpha}(\R,\M;0)\xibar_1^\dalpha
-2\sqrtwo\C_{\gamma\dalpha}(\R,\N;0)\xibar_2^\dalpha}
and likewise for $\Rtilde$. That the third argument of $\C$ is zero 
reflects the fact that  $\vev{\uq\,}=\vev{\uqtilde\,}
=0\,$; consequently $\Atot$ in \Cdef\ collapses to $\A_f.$
Given the transformation law \Rtrans, and  the constraints \constraints,
it is easily checked that $\delta S_\mass=0\,$.

Next we turn to the $\uQtilde\uPhi\uQ$ 
 contribution to the superpotential \superptl. Our
treatment of this term parallels the case of fundamental
hypermultiplets discussed in \dkmfour. As in that case, the key effect of this
coupling is to provide a fermion bilinear source term in the 
classical equation for $\uAbar\,$,
%\eqn\Abareqn{\D^2\uAbar\ =\ \sqrtwo i\,[\uchi\,,\uchitilde\,]\ ,}
namely $\D^2\uAbar = \sqrtwo i\,[\uchi,\uchitilde\,]\,,$
just like the $\sqrtwo i\,[\ulambda,\upsi]$ source term for $\uA$ 
itself \dkmone. This
observation is important because (as detailed in Secs.~4.3 and 7.4 of
\dkmone) 
the multi-instanton action \sinstfinal\
can actually be read off from the asymptotics
of the Higgs field, thanks to Gauss' law. In particular, the source term
for $\uA$ is responsible for the $\M\times\N$ fermion bilinear contributions
to $S_\inst^0$ 
given above. By identical arguments, the hypermultiplet source term 
%in \Abareqn\ 
for $\uAbar$
induces the following $\R\times\Rtilde$ bilinear contributions to the
$n$-instanton action:
\def\hyp{{\rm hyp}}
\eqn\Shypdef{S_\hyp\ =\
\ 8\pi^2\,\Tr_n\big(\A'_{f}(\R,\Rtilde)\,\Lambda\big)
-\  
4\sqrtwo\pi^2\,\rho_k^\alpha\,\vhiggsa{}_\alpha{}^\beta\,\rhotilde_{k\beta}\ .}
This is the same as the $\M\times\N$ bilinear in \sinstfinal, 
with $\M\rightarrow\R,$ $\N\rightarrow\Rtilde,$ and $\vbarhiggs\rightarrow
\vhiggs.$

\def\Squad{S_{\rm quad}}
\def\lilA{{\scriptscriptstyle A}}
\def\lilB{{\scriptscriptstyle B}}
\def\lilC{{\scriptscriptstyle C}}
\def\lilD{{\scriptscriptstyle D}}
Finally there are the fermion quadrilinear terms $\Squad$, 
consisting of one collective coordinate each drawn 
from $\M,$ $\N,$ $\R$ and $\Rtilde.$ Such terms are generated by each of the
six Yukawa couplings and each of the three
Higgs kinetic energy terms in the component
Lagrangian. As in \dkmfour,  $\Squad$
is fixed uniquely by the requirement that, when combined with 
 $S_\hyp,$ it forms 
a \susy\ invariant. In practice, we have derived $\Squad$ from an explicit
calculation in
the 2-instanton sector. Alternatively its construction is motivated by the
following arguments. By dimensional analysis, $\Squad$ depends neither on
the mass $m$ nor on the VEVs $\vhiggs$ or $\vbarhiggs.$ 
Consequently it must respect
the enlarged Lagrangian symmetries that emerge when $m,$ $\vhiggs$ 
and $\vbarhiggs\rightarrow0.$ First, 
as mentioned above, when $m\rightarrow0,$ the $N=2$ \susy\ is promoted
to $N=4$, so that $\M,$ $\N$, $\R$ and $\Rtilde$ must appear in $\Squad$
 in an $SU(4)_R$-invariant manner.  This is arranged by relabeling
$\{\M,\N,\R,\Rtilde\}\rightarrow\{\M^1,\M^2,\M^3,\M^4\}$
and forming $\Squad$ from the invariant tensor 
$\epsilon_{\lilA\lilB\lilC\lilD}$ acting on these indices. 
Second, in the limit that $m,$ $\vhiggs$ and $\vbarhiggs$ 
vanish, \susic\ invariance
is enlarged to superconformal invariance. Consequently $\Squad$ must 
vanish when
any of the $\M^\lilA_\gamma$ is proportional to a superconformal adjoint zero
 mode. These are the modes with $\M^\lilA_\gamma\propto a_{\gamma\dalpha}
\etabar_\lilA^\dalpha$ where $\etabar_\lilA$ is an arbitrary Grassmann 
parameter.\foot{This may be seen as follows \CGTone. In the
notation of \refs{\dkmone-\dkmfour}, the
supersymmetric adjoint zero modes  are
the ones with $\M^\lilA\propto b\xi_\lilA$, so that the superconformal modes
(which are obtained by promoting $\xi_\lilA
\rightarrow \xi_\lilA(x)=x\etabar_\lilA$)
 are $\propto b x\etabar_\lilA$. But since, in the explicit
expressions for the fermions,  $\M^\lilA$
is always left-multiplied by the quantity $\bar U$, and since
$\bar U\cdot(a+bx)=0,$ one can swap
$bx\etabar_\lilA$ for the space-time constant $-a\etabar_\lilA$.}
 Note that $\Lambda_f(\M^\lilA,\M^\lilB)$ vanishes in this
instance thanks to \constraints, so that superconformal invariance is 
guaranteed
if one builds $\Squad$ out of the $\Lambda_f(\M^\lilA,\M^\lilB)$ and/or the
$\A'_f(\M^\lilA,\M^\lilB)$ 
since $\bigL$ is generically invertible. In addition, 
one may guess
that $\Squad$ has the same structure as in the case of fundamental
 hypermultiplets, where it is proportional to $\Tr_n\,\A'_{f}(\M,\N)
\Lambda_{{\rm hyp}}$,
where $\Lambda_{\hyp}$ is an $n\times n$ antisymmetric matrix formed from the
collective coordinates of the fundamental fermions \dkmfour. 
Taken together, these
considerations motivate the result
\eqn\Squadef{\Squad\ =\ \pi^2\,\epsilon_{\lilA\lilB\lilC\lilD}
\,\Tr_n\,\A'_f(\M^\lilA,\M^\lilB)
\Lambda_f(\M^\lilC,\M^\lilD)\ .}
The overall constant is fixed by requiring that $S_\hyp+\Squad$ be invariant
under the supersymmetry transformations \susytrans\ and \Rtrans\ (checking
this is straightforward once one rewrites $\A'_f=\bigL^{-1}\cdot\Lambda_f$ 
and uses
$\delta\bigL^{-1}=-\bigL^{-1}\cdot\delta\bigL\cdot \bigL^{-1}$ together
with the self-adjointness of $\bigL^{-1}$). 

In summary, the $n$-instanton action for mass-deformed $N=4$ SYM theory is 
given by
\eqn\Stotdef{S_\inst^\tot\ =\ S_\inst^0+S_\mass+S_\hyp+\Squad\ .}

%\newsec{One-instanton calculation}
\bf3. One-instanton calculation. \rm
We now specialize to the 1-instanton sector, in which case $a,$ $\M,$
$\N,$ $\R$ and $\Rtilde$ reduce, respectively, to the $2\times1$ matrices
\eqn\listmat{
\Big({w\atop X}\Big)\ ,\ \
\Big({\mu\atop 4\xi_1}\Big)\ ,\ \
\Big({\nu\atop 4\xi_2}\Big)\ ,\ \
\Big({\rho\atop R}\Big)\ ,\ \
\Big({\rhotilde\atop \tilde R}\Big)\ .}
Also, since all ``$1\times 1$ antisymmetric matrices'' vanish by definition,
$S^\tot_\inst$ collapses to 
\eqn\oneinst{S^\tot_\inst(n=1)\ =\ 
{8\pi^2\over g^2} + 16\pi^2|\vhiggsa|^2|w|^2
+  4\sqrtwo\pi^2(\mu\barvhiggsa\nu-\rho\vhiggsa\rhotilde)
+2im\pi^2(2\rhotilde\rho+\tilde RR)\ .}
In previous work (Eq.~(21) of \dkmtwo\ and Eq.~(7.20) of \dkmfour)
 we derived an explicit
representation of the $n$-instanton contribution  to the
$N=2$ prepotential $\F(\vhiggs),$ as an integral over all $n$-instanton
supermoduli, excepting the four translational modes $d^4X$ and their
$N=2$ superpartners $d^2\xi_1$ and $d^2\xi_2$. In the present model, 
using these equations, together with
the 1-instanton Pauli-Villars (PV) measure given in \tHooft, one 
derives
\eqn\Fonedef{\eqalign{\F(\vhiggs)\Big|_{1\hbox{-}\rm inst}\ &=\
8\pi i
\int \eighth\pi^{-12}\,
d^4wd^2\mu d^2\nu d^2\rho d^2\rhotilde d^2R d^2\tilde R\,\exp\big(
-S^\tot_\inst(n=1)\big)\cr
&=\ -{i\over\pi}\Big({m^4\over2\vhiggs^2}-{m^2\over4}\Big)
e^{-8\pi^2/g^2}\ .}}
(As per \Seiberg\ this vanishes as $m\rightarrow0$ since the
 $R$ and $\tilde R$ modes are no longer lifted by the action \oneinst.)
Actually  the prepotential is only defined up to arbitrary
constant and linear terms, $\F\rightarrow\F+A\vhiggs+B$. Henceforth
we set $A=B=0\,$; as shown in \refs{\Fucito,\dkmtwo},
this is the unique choice for which $\F$ and the quantum
modulus $u$ are simply proportional to one another by Matone's 
relation \refs{\MAtone,\BMSTY},
order by order in the instanton expansion:\foot{The all-instanton-orders 
proof of  Matone's relation given in \dkmtwo, while
ostensibly limited to  pure $N=2$ SYM theory, applies equally
in the presence of fundamental or adjoint hypermultiplets.}
\eqn\matonerel{u(\vhiggs)\Big|_{n\hbox{-}\rm inst}\ =\
2\pi in\,\F(\vhiggs)\Big|_{n\hbox{-}\rm inst}\ ,\qquad n>0\ .}

As a consistency check on Eq.~\Fonedef, we consider the RG decoupling
limit \refs{\SWtwo,\FPone,\dkmfive} 
in which the hypermultiplet becomes infinitely massive, and the
model reduces to pure $N=2$ SYM theory:
\def\PV{{\scriptscriptstyle\rm PV}}
\eqn\RG{m\rightarrow\infty\ , \qquad g\rightarrow0\ ,
\qquad m^4 e^{-8\pi^2/g^2}=\Lambda_\PV^4\ .}
In this  limit Eq.~\Fonedef\ reduces to
 $-i\Lambda_\PV^4/2\pi\vhiggs^2$, which is indeed the correct
1-instanton contribution to the prepotential for $N=2$ SYM theory
\FPone, in terms of the dynamically generated scale $\Lambda_\PV$ in the
PV scheme. 

In what follows we will focus particularly on the $\CO(m^2)$ term
in \Fonedef, where we will claim a discrepancy with the predictions
of Seiberg and Witten \SWtwo. Note that
 this term  does not survive the double scaling limit \RG. Nor does it
contribute to the low-energy effective $U(1)$ Lagrangian, which depends
on the prepotential only through its
derivatives $\F''(\vhiggs),$ $\F'''(\vhiggs),$ and $\F''''(\vhiggs).$
It does, however, contribute to $u$; Eq.~\matonerel\ gives
\eqn\uinstdef{u(\vhiggs)\Big|_{1\hbox{-}\rm inst}\ =
2\pi i\,\F(\vhiggs)\Big|_{1\hbox{-}\rm inst}\ =\
\Big({m^4\over\vhiggs^2}-{m^2\over2}\Big)
e^{-8\pi^2/g^2}\ .}
Now we compare this first-principles semiclassical
result with the predictions of \SWtwo.

%\newsec{Predictions of Seiberg and Witten}
\bf4. Predictions of Seiberg and Witten. \rm
As proposed in \SWtwo, the exact solution of the mass-deformed $N=4$ model
is governed by the elliptic curve
\def\E{{\cal E}}
\eqn\curve{y^2\ =\ (x-\E_1)(x-\E_2)(x-\E_3)\ ,\quad
\E_i=e_i\utilde+\quarter e_i^2m^2}
where the $e_i$ are the modular forms
\eqn\eidef{\eqalign{e_1\ &=\ \textstyle{2\over3}+16q+16q^2+\CO(q^3)\cr
e_{\left\{{2\atop3}\right.}
\ &=\ -\textstyle{1\over3}\,\mp\, 8q^{1/2}-8q\,\mp\, 32q^{3/2}-8q^2\,\mp\,
48q^{5/2}+\CO(q^{3})}}
Here $q=\exp(2\pi i\tau)$ 
is the 1-instanton factor, with $\tau$ the complexified
coupling $\tau={\theta\over2\pi}+{4\pi i\over g^2}$. 
It is desirable to express the curve parameter $\utilde$ in terms of the
physical condensate $u = \langle\Tr\uA^2\rangle$. To this end we examine
the singularities of \curve, which occur when two of the $\E_i$ coincide.
Since $\sum e_i=0$ these lie at
\eqn\coincc{\utilde\ =\ \big\{\quarter e_1m^2\ ,\ 
\quarter e_2m^2\ ,\ \quarter e_3m^2\big\}\ .}
On the other hand, we presume to know the singularities in the $u$ plane
as well, at least in the RG decoupling limit \RG. There is, first, the
``semiclassical'' singularity at $u\simeq m^2/4$ where a component of
the hypermultiplet becomes massless (see Eq.~\superptl\ \it ff.\rm); the
remaining ``strong coupling'' singularities are those of the pure
$N=2$ SYM theory \SWone, which lie at\foot{We use the fact \FPone\ that
the dynamical scale $\Lambda$ used in \SWone\ is related to
$\Lambda_\PV$ by $\Lambda^4=4\Lambda^4_\PV.$} $u\simeq\pm2\Lambda_\PV^2.$
Thus
\eqn\coincd{u\ \simeq\ \big\{
\textstyle{1\over4}m^2\ ,\ -2\Lambda_\PV^2\ ,\ 2\Lambda_\PV^2\big\}\ .}
Comparing Eqs.~\coincc\ and \coincd\ and using Eqs.~\RG\ and \eidef\
leads to 
\eqn\dicta{\utilde\ =\ u-m^2\,\big(\textstyle{1\over12}+\sum_{n=1}^\infty
\alpha_n q^n\,\big)}
where the $n$-instanton coefficients 
$\alpha_n$ are not determined by the above argument.  Note that in the massless
limit we recover simply  $\utilde=u$ (in contrast to the $N_F=4$ model
where there is a finite multi-instanton renormalization
\refs{\HS,\dkmfive}). With 
%the substitution 
Eq.~\dicta, it is easily verified that the curve \curve\ reduces in the 
limit \RG\ to that of the $N=2$ SYM model, namely \SWone:
\eqn\curveSYM{y^2\ =\ (x-u)(x^2-4\Lambda_\PV^4)\ .}
To see this, one needs to shift  $x$ as follows, prior to taking the limit:
\eqn\xshift{x\ \rightarrow\ x-\textstyle{1\over3}u
+{1\over18}m^2+\CO(q)\ .}

The above discussion follows closely that of Sec.~16.2 of \SWtwo. However,
these authors claim a more detailed knowledge of the ``dictionary'' between
$\utilde$ and $u$, namely
\eqn\dictb{\utilde\ =\ u-\textstyle{1\over8}e_1 m^2\ .}
This is a special case of \dicta\ containing precise predictions for all
the $\alpha_n$, e.g., $\alpha_1=\alpha_2=2.$
 We will soon find that Eq.~\dictb\ is mistaken starting
at the 1-instanton level.\foot{The reader can check (see footnote 7 of
\SWtwo) that modifying
Eq.~\dictb\ to the more general relation \dicta\ does not affect
the  residue conditions developed in Secs.~15 and 17 of$\,$ \SWtwo.
In addition to \dictb, Seiberg and Witten also employ a more precise
version of the shift \xshift, but the variable $x$ is just a dummy of
integration and carries no physical meaning, at least in 4 dimensions.}
To see this, one extracts $\partial\vhiggs/\partial\utilde$ in the
usual way as a period of the curve:\foot{The first equality here
 assumes that $q^{1/2}m^2 \ll u \ll m^2,$ where $q$ is real, positive, and
$\ll1,$ from which it follows that $\E_1>\E_2>\E_3$.
This regime includes the RG limit \RG,
but is  more general since it does not assume $m\rightarrow\infty.$
 In contrast, in the chiral limit $m\rightarrow0$ (still
with $q$ small), the
$\E_i$ collapse to  $e_iu$ and the ordering switches to
$\E_1>\E_3>\E_2$. Nevertheless, 
since the only difference between the two orderings 
is which branch of $\sqrt{q}$ is taken, the
formulas that follow, which depend only on integer powers of $q$, are 
 valid in either regime.}
\eqn\period{\eqalign{\sqrtwo\,{\partial\vhiggs\over\partial\utilde}\,
&=\, {2\,K(k)\over\pi\,\sqrt{\E_1-\E_3}}\cr&=\ \textstyle
z+(m^2z^{3}+{\textstyle{3\over4}} m^4z^{5})q+(3m^2z^{3}+6m^4z^{5}
+{105\over64}m^8z^{9})q^2+\CO(q^3) ,}}
where $K(k)$ is an elliptic function of the first kind, 
%and
%\eqn\kdef{k\,=\,\Big({\E_2-\E_3\over\E_1-\E_3}\Big)^{1/2}\ \ ,\qquad
%z\,=\,\big(\utilde+\twelfth m^2\big)^{-1/2}\ .}
$k=\big[(\E_2-\E_3)/(\E_1-\E_3)\big]^{1/2}$, and
$z=\big(\utilde+\twelfth m^2\big)^{-1/2}$.
Inverting \period\ gives
\eqn\invert{\utilde\ =\ -\twelfth m^2+\hf\vhiggs^2+
\big(2m^2+{m^4\over\vhiggs^2}\big)q +
\big(6m^2+6{m^4\over\vhiggs^2}-6{m^6\over\vhiggs^4}
+{{5\over4}}{m^8\over\vhiggs^6}\big)q^2 \,+\,\CO(q^3)}
which combines with \dicta\ to give, finally,
\eqn\useries{u\ =\ \hf\vhiggs^2+
\big((2+\alpha_1)m^2+{m^4\over\vhiggs^2}\big)q +
\big((6+\alpha_2)m^2+6{m^4\over\vhiggs^2}-6{m^6\over\vhiggs^4}
+{{5\over4}}{m^8\over\vhiggs^6}\big)q^2 \,+\,\CO(q^3)\ .}
By comparing \useries\ with the explicit 1-instanton calculation
\uinstdef\ we see that $\alpha_1=-5/2$, contradicting the Seiberg-Witten
prediction $\alpha_1=2$ from Eq.~\dictb. We expect the higher $\alpha_n$
to differ as well.

%\newsec{Discussion}
\bf 5. Discussion. \rm
We stress that the $\alpha_n$ are  not  predictions contained in
 the curve itself. Certainly they are in no way tests of the modular
properties of Eq.~\curve. They are merely
parameters in the ``dictionary'' \dicta\ between $\utilde$ and $u$. The
fact that the Seiberg-Witten proposal for this dictionary, Eq.~\dictb,
needs to be modified, is similar to  the $N_F=3$
\refs{\Aoyama,\HS} and $N_F=4$ \refs{\dkmfour,\dkmfive} models with
fundamental hypermultiplets. In the $N_F=3$ theory, however, only
a single constant needs to be calculated (namely $\alpha_2$), whereas
in the $N_F=4$ model, as in the present model, an infinite number of instanton
orders contribute. As mentioned earlier the $N_F=4$ model has an additional
set of finite instanton renormalizations even in the massless limit 
\refs{\dkmfour,\dkmfive} which are necessarily absent in the $N=4$ model
\Seiberg.
%The $N_F=4$ model has an additional complication:
%even when all four ``quark'' masses $m_i$ are zero, an 
%all-even-instanton-orders
%finite renormalization distinguishes the microscopic $SU(2)$ coupling
%$\tau$ from the  effective $U(1)$ coupling $\taueff(m_i=0)$; contrary to
%\SWtwo\ it is the
%latter quantity that should appear in the massive curve \dkmfive. In contrast,
%in the present model, the enlarged $N=4$ \susy\ that emerges when
%$m\rightarrow0$ protects against such a renormalization \Seiberg; 
%extra unbroken fermion zero modes appear that can no longer be saturated 
%with a %four-antifermion correlator (say). 

Aside from the modified dictionary between $\utilde$ and $u$,
 it is natural  to assume that
the solution of the mass-deformed $N=4$
model is still given by the curve \curve. This is the same
``minimal assumption'' 
%that is 
made in \HS\ and \dkmfive\ for the $N_F=3$
and $N_F=4$ models, respectively. We underscore that this is an
\it assumption\rm, which ought to be tested against a
first-principles instanton calculation. As explained in Sec.~1 of
\dkmfive, for the $N_F=3,4$ models the first such nontrivial tests
lie in the 3-instanton sector, whose moduli space is quite complicated.
Fortunately, in the mass-deformed $N=4$
theory, such  tests are available at the 2-instanton
level. To see this,  consider the $\CO(q^2)$
contribution to $u$ in Eq.~\useries. There are four terms at this
level, proportional to
$m^2,$ $m^4,$ $m^6$ and $m^8.$ The  $m^2$ piece is, again, an
entry in the ``dictionary'' \dicta. Also the  $m^8$ piece is
not an independent prediction; its coefficient is fixed by the
RG decoupling relation \RG, which 
is built into the curve as well as into the instanton calculus 
(indeed  it is consistent with 
 $\F\big|^{}_{2\hbox{-}\rm inst}=(4\pi i)^{-1}u\big|^{}_{2\hbox{-}\rm inst}=
-5i\Lambda_\PV^8/16\pi\vhiggs^6$ 
in the $N=2$ SYM model \dkmone). However,
the  $m^4$ and $m^6$ coefficients constitute two \it bona fide \rm testable
predictions of the curve \curve, in particular of its elegant
modular properties.
As such, they are quite important.  In a forthcoming paper \dkmseven\ we shall
compare these coefficients against the results of an explicit 2-instanton
calculation.
%$$\scriptscriptstyle{**************************************}$$

 We acknowledge a clarifying discussion with Frank Ferrari about the dyon
spectrum.

\listrefs
\bye